# Large-scale epitaxial integration of single crystalline BiSb topological insulator on GaAs (111)A

*Mohamed Ali Khaled[1]\*, Leonardo Cancellara[1], Salima Fekraoui[1], Richard Daubriac[1], François Bertran[2], Chiara Bigi[2], Quentin Gravelier[1], Richard Monflier[1], Alexandre Arnoult[1], Corentin Durand[1], Sébastien R. Plissard[1]\**

AUTHOR ADDRESS

[1] LAAS-CNRS, Université de Toulouse, 31400 Toulouse, France.

[2] Synchrotron SOLEIL, BP48, L'Orme des Merisiers, Saint-Aubin, 91192 Gif-sur-Yvette, France.

ABSTRACT.

Topological insulators (TI) are promising materials for future spintronics applications and their epitaxial integration would allow the realization of new hybrid interfaces. As the first materials studied, Bismuth Antimony alloys ($Bi_{1-x}Sb_x$) show great potential due to their tuneable electronic band structure and efficient charge-to-spin conversion. Here, we report the growth of $Bi_{1-x}Sb_x$ thin films on GaAs (111)A substrates following two different protocols. For the conventional epitaxy process, the grown films show excellent crystallinity and twin domains corresponding to an in-plane 180° rotation of the crystalline structure. Domain walls are found to be composition-dependent and have a lower density for Antimony-rich films. For the optimized process, depositing an Antimony bilayer prior to BiSb growth allows achieving single crystallinity of the TI films. The topologically protected surface states are evidenced by ex-situ ARPES measurements for domains-free and conventional films. To the best of our knowledge, this work presents the first large-scale epitaxial integration of single crystalline $Bi_{1-x}Sb_x$ thin films on industrial substrates.

## 1. Introduction

Three-dimensional Topological insulators (3D TI) are materials with semiconducting bulk behaviour and gapless metallic states at their boundaries.[1] As predicted by Kane and Mele, in a material with strong spin orbits coupling, breaking the space inversion symmetry while preserving the time-reversal one, generates topologically protected surfaces states (TSS) displaying spin-momentum locking.[1,2] Moreover, once an electric current is applied through the material, a spin accumulation can be observed at the material surface.[2–4] This effect could be compared to those observed for spin-Hall effect in heavy metals and Rashba-Edelstein effect in ferromagnets.[5] Therewith, the induced metallic surface states, in TI materials, exhibit a robust spin polarisation which enables an efficient charge-to-spin conversion.[4,5] Finally, the transfer of spin's angular momentum from the TI's surface enables magnetization switching in an adjacent magnetic layer having a perpendicular anisotropy (PMA).[6,7] Such phenomenon is referred to as current-induced spin-orbit torques (SOT), which allows a deterministic manipulation of magnetization in a magnetic memory device.[8,9]

In this context, Bismuth Antimony ($Bi_{1-x}Sb_x$) is one of the most promising materials for Spintronics applications when its Sb composition lies between 7 and 22%, which corresponds to its semiconductor window.[10–12] Non-trivial TSS have already been reported by Hsieh *et al.* for a sample with x=0.1, in which an odd fermi level band crossing is observed by ARPES measurements.[11] Since this pioneer work, similar properties have been observed for multiple materials ($Bi_2Se_3$, $Bi_2Te_2Se$, $BiSe_2Te$, BiSbTeSe alloys, …) creating a plethora of candidates for future applications.[13–16] Considering SOT-MRAM applications, BiSb alloys outstand other TIs thanks to their large spin-hall angle ($\theta_{SH} \approx 52$) generating a high spin-orbit field, which leads to a very low switching current for the magnetization of an adjacent ferromagnetic layer.[17] The origin of such unique properties remains unclear mostly due to the complex transport mechanisms in BiSb. Indeed, different carrier channels have a contribution on the global electrical transport such as the TSS, the thermally activated carriers through the narrow band gap, and, possibly, the defects-enabled leakage currents. The surface and bulk conduction channels are expected to be tuneable by changing the composition, whereas the latter is directly linked to the material synthesis.[18]

This driving search for efficient TI materials led to multiple attempts for integrating BiSb thin films on industrial substrates. Recently, Yao *et al*. reported the epitaxial growth of BiSb thin films by Molecular Beam Epitaxy (MBE), on GaAs (111)B substrates.[19] Yet, the obtained films suffer from high surface roughness and a composition out of the TI range, which is detrimental for device integration. Similar difficulties in controlling the composition appears at high growth temperature (250°C) on $BaF_2$ (111) substrates.[20] Finally, the BiSb growth on sapphire by sputtering techniques leads to polycrystalline films above 50nm.[21] Smooth MBE films with good crystalline quality were obtained on GaMnAs-buffered GaAs (001) substrates.[22] Moreover, the difficult integration of a buried MRAM stacking before TI growth motivated the search for different substrates and growth strategies.[23,24] In previous studies, we reported the 2D MBE growth of BiSb films with TSS on GaAs (111)A and (001).[25,26] Room-temperature Hall measurements revealed a semiconducting behavior of the bulk, transiting to a metallic one below 100K. Typical hole Hall mobilities at 20K are 1430 and 7620 $cm^2/(V·s)$ for 450 nm and 1 μm thick films, respectively.[25,26] For the integration of such material in future low-energy consumption spintronic devices, two essential points need to be considered: the TI films' uniformity and surface smoothness need to be optimized for magnetic material deposition along with preserved topological surface states are required.

In this study, we report the successful epitaxial growth of TI $Bi_{1-x}Sb_x$ films on (111)A oriented GaAs substrates by MBE. We study the influence of the Antimony composition, within the TI range, on the films' crystallinity and the microstructure morphology for a series of 2µm-thick samples. In-depth X-ray diffraction (XRD) analysis and transmission electron microscopy (TEM) observations are performed to unveil the nature of the observed domains. TSS features are confirmed by ARPES measurements since the surface bands are crossed an odd number of times at the Fermi level. A new growth protocol is proposed enabling the epitaxial growth of single-crystalline BiSb films on 2 inches wafers.

## 2. Results and discussion

### 2.1. Growth protocol

$Bi_{1-x}Sb_x$ thin films (x = [0.7; 0.11; 0.15; 0.19]) are grown on (111)A oriented GaAs substrates having a 2° miscut in the [1-10] direction by molecular beam epitaxy (MBE). The surface reconstruction is measured during growth by reflection high energy electron diffraction (RHEED). 2 inches GaAs wafers from AXT are loaded in our MBE system and transferred into the preparation chamber to be degassed at 300°C for 1 hour. Wafers are then transferred into the growth reactor and heated up to the deoxidation temperature of 635°C under $2x10^{-5}$ Torr of Arsenic as illustrated in **(Figure 1.a)**. As soon as native oxide removal is confirmed by RHEED observations, 300 nm of GaAs is homoepitaxially grown at 580°C in one hour. Finally, after reaching 160°C, the $Bi_{1-x}Sb_x$ growth is initiated by opening simultaneously the Bismuth and Antimony shutters. The BiSb growth rate is set to 260 nm per hour. Note that the Arsenic flux used for deoxidation is kept down to 400°C.

### 2.2. Material characterizations

**(Figure 1.c)** shows a 30° tilted SEM image of an as-grown $Bi_{0.89}Sb_{0.11}$ 300nm thick thin film. Randomly distributed grains can be observed (bright areas) within an overall uniform layer (darker area). A zoom-in is proposed in **(Figure 1.d)**, enlightening a rather smooth layer-by-layer growth of the continuous matrix, which opposes to the rough grain surface. Such patterns are found to be less visible for higher thicknesses, while the domain structure is found to be persistent as presented in **(Figure S.1)** for a set of 2 µm thick films with different compositions. On the other hand, the domain walls density is composition-dependent and decreases when the Antimony amount is increased as shown in **(Figure S.2)**. Note also that triangular shapes can be observed, that have all the same orientation, which is characteristic of an epitaxial growth.

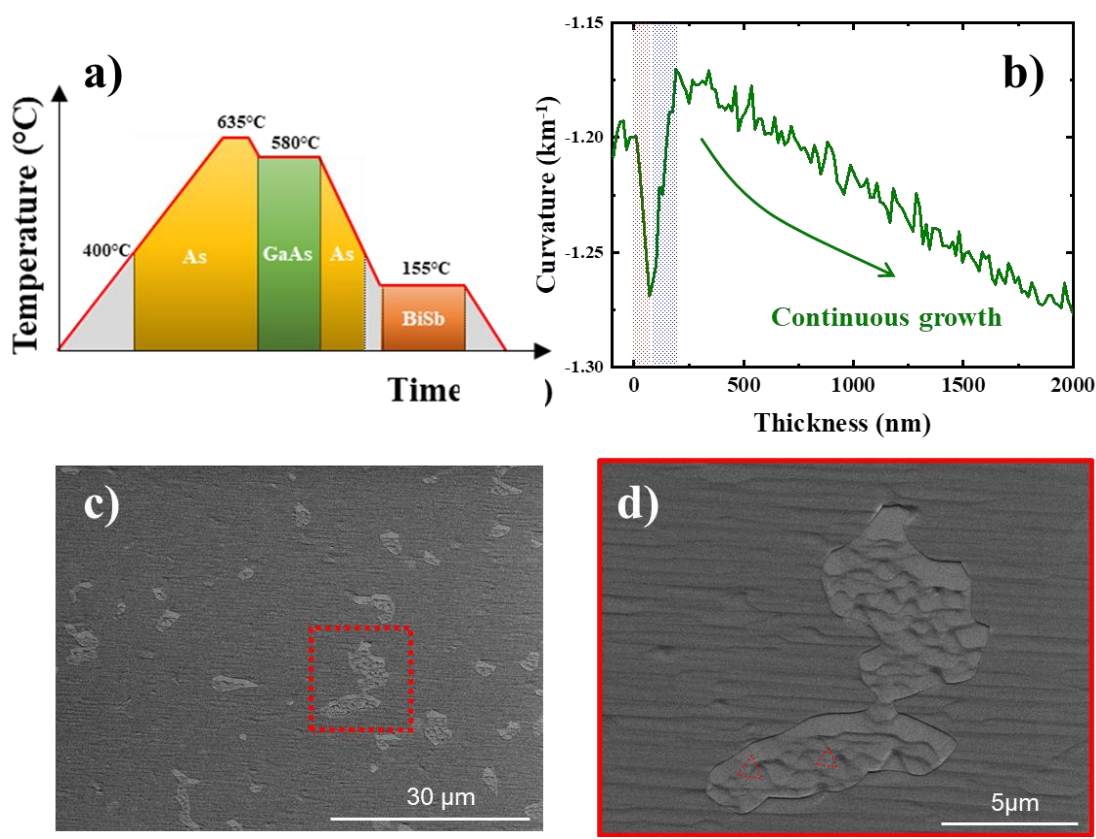

**Figure 1. a)** Temperature vs time illustrative diagram of the MBE growth process of our $Bi_{1-x}Sb_x$ thin films. **b)** Curvature as a function of growth time. Different growth steps are indicated. Note that Bi and Sb shutters are opened and closed simultaneously. **c)** Surface morphology of the as-grown BiSb layer probed by SEM. The zoomed-in area indicated by a red square is shown in **(d)**. Target composition and thickness are x=0.11 and 300nm.

Moreover, we observe that the film morphology depends on the $Bi_{1-x}Sb_x$ layer thickness. Indeed, holes are present for thicknesses below 300nm, and domains emerge beyond 50nm. The surface curvature presented in **(Figure 1.b)**, and measured during growth shows 3 different regimes. A drop of the curvature in the first 30 nm is observed (red filling), indicating the nucleation of the initial islands. Next, a jump in the curvature measurement is observed (blue area) indicating the coalescence of the initial islands. Beyond 200 nm, a continuous strain accumulation is observed until the end of the growth. In order to study the films' crystallinity, X-ray diffraction analysis in a $\theta$-$2\theta$ geometry is performed and presented in **(Figure 2.a)** for the set of 2µm thick samples and for the first three orders of diffraction. The absence of any parasitic phase inside the BiSb films is confirmed by the well-shaped single diffraction peaks at each measured order. Also, the diffraction analysis reveals a BiSb(0001)/GaAs(111)A epitaxial relationship between the grown films and the substrate. Note that the 3 low intensity peaks observed near $2\theta = 42°$, 54°, and 64° are generated by the experimental setup and are not related to the BiSb film. Zoomed-in $\theta$-$2\theta$ diagrams around the angular region corresponding to the second order of diffraction are presented in **(Figure 2.b)**. Contrary to the GaAs peaks (see dashed lines in **(Figure 2.a)**), a clear peak shifting to higher angles is observed for BiSb (0006) peaks when increasing the Antimony composition. Such a trend reveals the expected evolution of the out-of-plane lattice parameter as a function of the composition. Interestingly, unfamiliar peaks' intensity changes are observed depending on both the azimuthal angle ($\Phi$) and whether alignments are performed along the substrate or the film. In order to understand and identify the origin of such behaviour, reciprocal space maps

(RSM) are collected around the $Bi_{0.89}Sb_{0.11}$(0006) diffraction peak at different azimuthal angle Φ (in-plane sample rotation around the z-axis). RSM performed at Φ=0°,60°,180°, and 270° are shown in **(Figure 2.c-f), respectively**. Different observations can be made from these symmetric Bragg reflections (depending only on the out-of-plane lattice component).

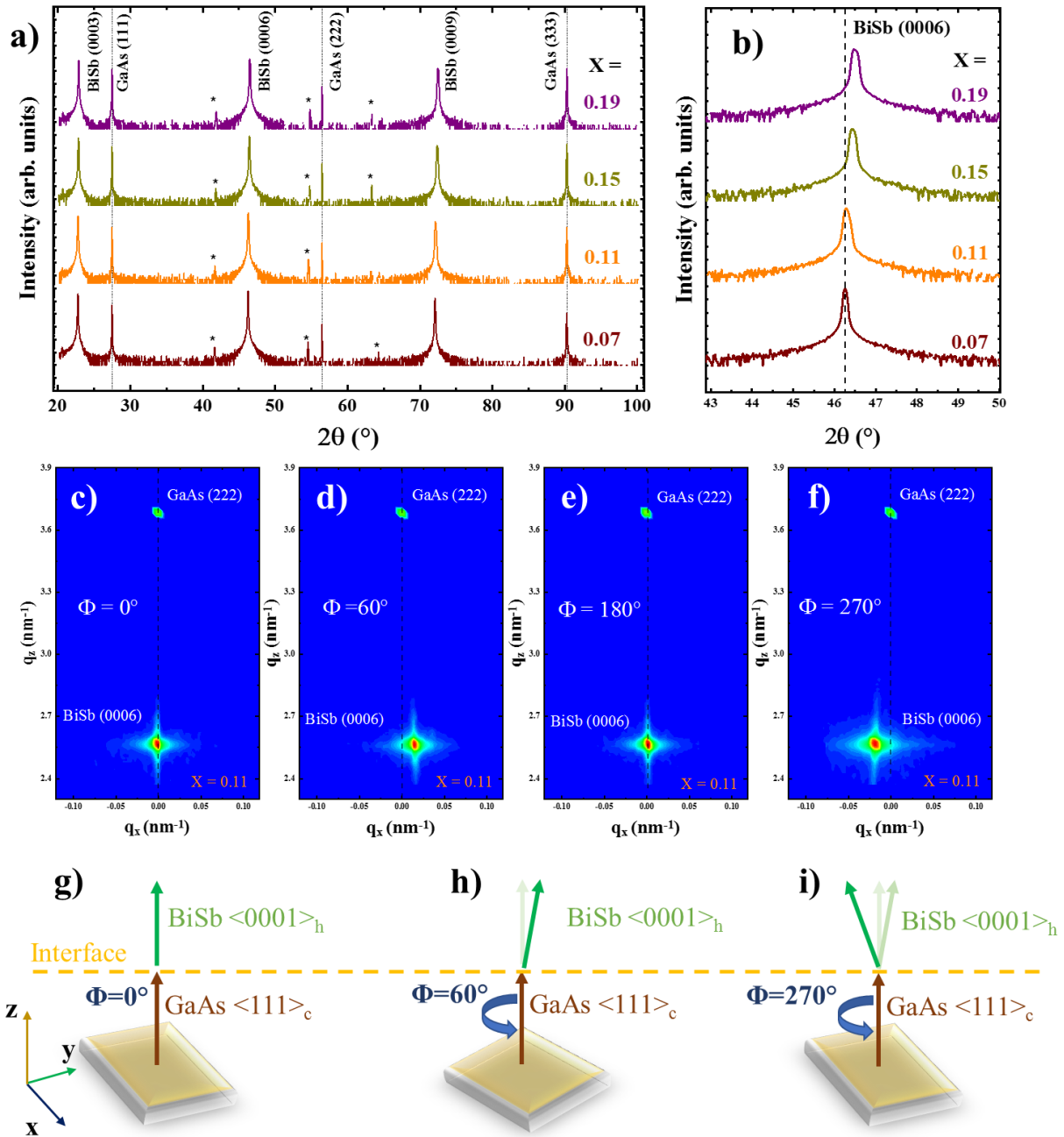

**Figure 2.** X-ray diffraction analysis of four $Bi_{1-x}Sb_x$ thin films with x = [0.7; 0.11; 0.15 0.19]. **a)** θ-2θ diffraction patterns up to the third diffraction order. Peaks indicated by asterisks (*)

are generated by the sample holder. **b)** Zoomed-in region at an angular range around the BiSb (0006) reflection. The dashed line is a guide for the eyes. The peaks shift to higher angles when increasing the Antimony composition. Reciprocal space mapping around the $Bi_{0.89}Sb_{0.11}$ (0006) reflection and the GaAs (222) reflection are presented for azimuthal angles of **(c)** $\Phi=0°$, **(d)** $\Phi=60°$, **(e)** $\Phi=180°$, and **(f)** $\Phi=270°$. The BiSb (0006) reflection shifts along $q_x$ when rotating the sample around $\Phi$. **(g)**, **(h)** and **(i)** Schematics of the real-space crystallographic-axis relation at the film-substrate interface for $\Phi$ azimuthal angles of 0°, 60° and 270°, respectively. The $\Phi$ origin is arbitrarily defined for a condition where the BiSb(0006) and GaAs(222) reflections are axis-aligned with $q_x=0$ nm$^{-1}$.

As shown in **(Figure 2.c)**, two peaks are visible for different $q_z$ positions, which corresponds to both the BiSb (0006) film and the GaAs (222) substrate. As expected, the GaAs peak is narrower than the BiSb one, and no trail is observed between both peaks, which is compatible with a higher mosaicity of the BiSb layer compared to the monocrystalline substrate and a relaxation of the accumulated strain by misfit dislocations at the layer-substrate interface, as observed previously.[26] Yet, the full widths at half maximum of the rocking curves peaks are equal to 0.067°; 0.092°; 0.093°; 0.102° for x= 0.07; 0.11; 0.15; 0.19, respectively (See **(Figure S.3)**). These values are an order of magnitude lower than the ones recently reported for $Sb_2Te_3$ films grown on $Bi_2Te_3$-buffered Si (111) substrates[27], which confirms the high crystalline quality of the TI layer. Next, we measured the same RSM at $\Phi=60°$ (see **(Figure 2.d)**), and observed a clear shift towards positive $q_x$ values of the film's reflection. This shift is $\Phi$ dependent, with the highest value observed near 90° (See **(Figure S.4)**). When increasing the azimuthal angle further, the BiSb reflection crosses the $q_x=0$ axis at $\Phi=180°$ and drifts towards negative $q_x$ values ($\Phi=270°$). **(Figure 2.e.f)**. Such a behaviour is explained by the tilted orientation of the $<0001>_h$ axis of BiSb film compared to the $<111>_c$ one of the GaAs substrate as illustrated in **(Figure 2.g-i)**. Indeed, when the incident beam and tilting plane are parallel (Φ=0° and Φ=180°), the effect of the tilt is screened and the reticular planes of the film and the substrate seem to be perfectly parallel. Once the sample is rotated around the z-axis, the tilt angle is observable through the BiSb(0006) reflection shifting, and reaches a maximum value when the incident beam and the tilting plane are perpendicular (Φ=90° and Φ=270°). Thanks to these measurements, we estimated the tilt angle to be close to ≈1.0° independently of the sample composition. Such a small value is not expected to affect the films' properties.

The presence of different domains can have a crucial influence on the transport properties and the study of the interface is fundamental for future industrial integration. To access the local structure across these domain walls, we conducted high-resolution (S)TEM (HR-(S)TEM) measurements as shown in **(Figure 3.a)**, where both the GaAs and BiSb layers are identified.

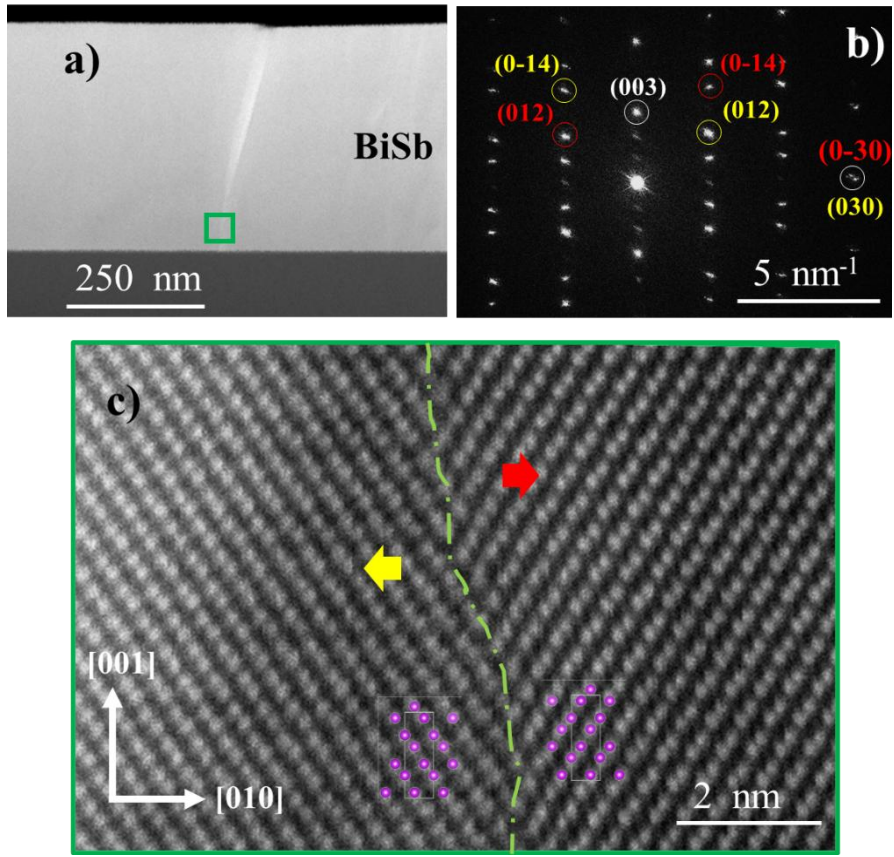

**Figure 3. a)** Cross-sectional STEM-HAADF image of a 450 nm-thick BiSb film grown on GaAs (111)A substrate. The grain boundary originates at the interface and reaches the film's surface. **b)** Corresponding fast Fourier transform (FFT) showing the diffraction pattern of the two BiSb grains in the <100> zone axis. Red and yellow annotations indicate the two different grains. **c)** High-angle annular dark-field (HAADF) scanning TEM image acquired in the green area, showing atomic arrangement within both domains' boundaries. The respective crystalline structures on both sides of the grain boundary are indicated in purple.

Two different grains can be observed separated by a slightly tilted boundary. The corresponding FFT shows equivalent but inverted in-plane crystal structures for the two domains. Reflections in red and yellow are equivalent to the same (*hkl*) families but appear at different reciprocal space positions due to an in-plane rotation of the structure (see (**Figure 3.b**)). Hints about the nature of such effect are visible on the pure in-plane reflections showing an intensity splitting as we can see for the (030) and (0-30) planes. On the other hand, the out-of-plane crystalline component exhibits no change across the grain boundaries. Note that extra weak reflections can be observed along the (001) direction that are measurement artefacts due to the presence of a Moiré pattern where the grains overlap. An STEM atomic-scale image of both grains is shown in **(Figure 3.c).** Their common boundary corresponds to the abrupt change in the crystal structure orientation, which is highlighted by the green dotted line. Left and right crystalline structures are pointed by yellow and red arrow, respectively. These different areas are responsible for the two reflection families observed in the FFT pattern. Illustrations of two different in-plane views of the same hexagonal structure of BiSb are indexed and perfectly fit the atomic positions in the STEM image. Hence, the observed domains correspond to the same BiSb crystalline structure with a different in-plane orientation. Such a texture is confirmed at the macroscopic scale by XRD pole figures acquired while performing complete rotations of the Φ-angle for different ψ angles (ranging from 0° to 90°) around the $Bi_{0.89}Sb_{0.11}$ (0009) Bragg reflection for a 300 nm-thick sample. Such thickness is ideal to collect an acceptable signal from the substrate while maximizing the one from the film. Both substrate and film exhibit a threefold symmetry as expected for R-3m crystalline structure along the trigonal direction. The substrate's reflections labelled in white in **(Figure 4.a.b)** are identified by taking a pole figure on a epi-ready GaAs (111)A wafer in the same previously used Bragg positions.

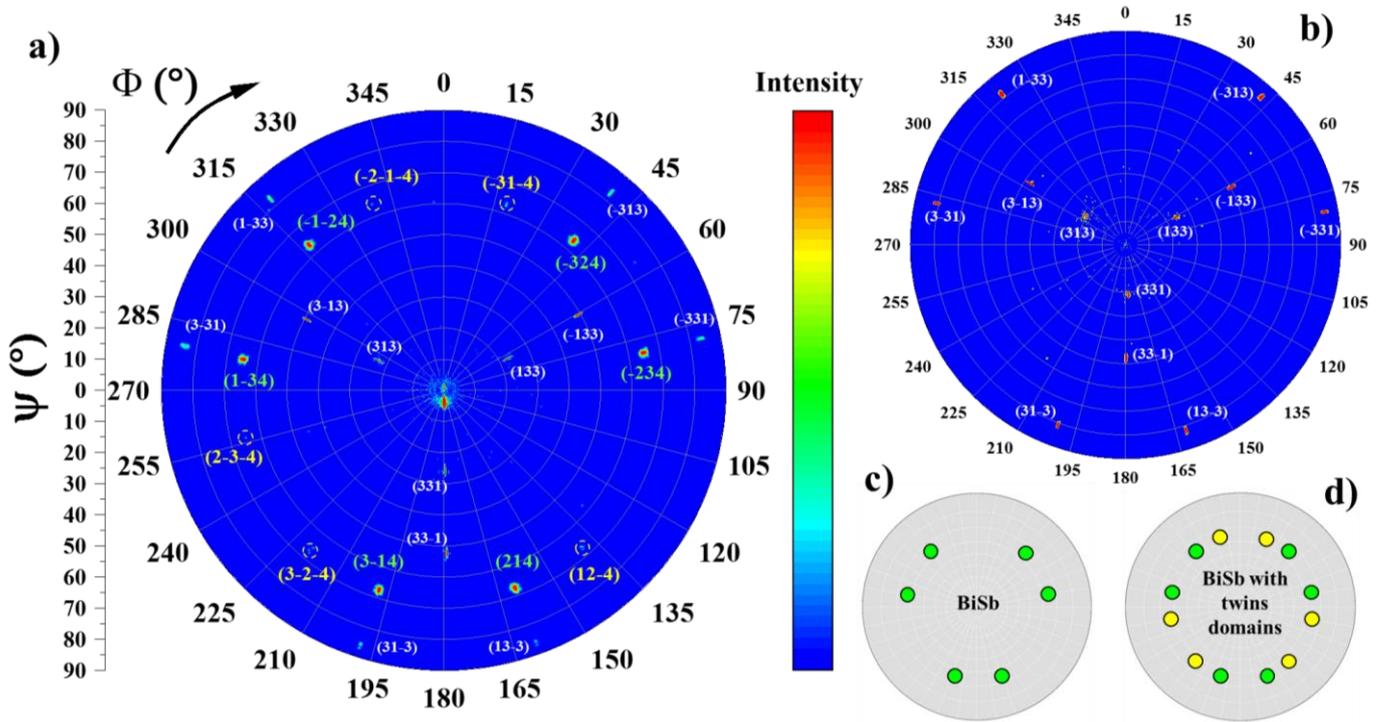

**Figure 4.** X-ray diffraction pole figures performed around the Bi$_{0.89}$Sb$_{0.11}$ (0009) Bragg reflection (2θ ≈ 72.2°) of a 300 nm-thick film **(a)**, and on a GaAs (111)A substrate **(b)**. The diffracted intensities are presented in a logarithmic scale. The labels near each reflection correspond to the attributed (*hkl*) planes. A specific color code is used for each type of reflection: white for the cubic GaAs substrate, green and yellow for both trigonal BiSb twins, using a hexagonal (*hkl*) system. Models obtained using WinWulff software for single crystalline and twinned BiSb films are presented in **c)** and **d)** respectively.[34]

WinWulff simulations are conducted to identify the BiSb film's reflections for two different systems: a BiSb layer without and with twin domains, as presented in **(Figure 4.c)** and **(Figure 4.d)**, respectively. For the former, six intense reflections are observed for the BiSb layer {214}, {-234}, {-324}, {-1-24}, {1-34}, and {3-14}, as labelled in green and are aligned with the substrate's reflections. Such an alignment points-out the epitaxial relationship between the GaAs and BiSb through the interface and during the growth process as reported previously. Moreover, weak extra reflections are observed (highlighted by yellow circles) at the same ψ angular positions of the BiSb spots but with an 180° in-plane φ rotation. As shown in the simulation presented in **(Figure 4.d)**, such reflections correspond to the same, yet rotated, BiSb crystalline structure and are indexed as {3-2-4}, {2-34}, {-2-1-4}, {-31-4}, and {12-4}. Those results are in perfect agreement with both HR-TEM and SEM observations in which the same domains' nature is proposed. We mention that different domain structures are already reported for BiSb films integrated on GaAs (001) substrates where a large out-of-plane axis tilting is observed. Such a difference is related to the different growth modes for each substrate as theoretically explained and experimentally confirmed in a recent study.[28] Another notable trend is the composition dependency of the domain walls density and the decreasing of the grain boundaries' length with the Sb concentration increasing, as shown in **(Figure S.2)**. This could be explained by the favorable Sb atoms bonding to the As-rich surface compared to Bi ones, creating a wetting layer and resulting in a layer-by-layer growth. Indeed, ionic Sb-As bounds are more stable and easier to create than Bi-As ones.[28] Consequently, the deposition of a Sb bilayer, before BiSb film growth, could affect the domain walls formations and promote the BiSb single crystalline growth. . In order to probe this hypothesis, a 500nm-thick BiSb film is grown after deposition of two Sb monolayers on GaAs (111) A and the surface morphology is observed by SEM **(see Figure 5.b)**. Contrary to the conventional process (see **(Figure 5.a)**), initiating the TI growth with the Sb-bilayer allows the nucleation of monocrystalline BiSb films with no domain structure. Instead, multiple triangular-shaped defects are observed on the film's surface. Such structural features seem to be randomly distributed, while always pointing in the same in-plane direction hinting at the epitaxial relation between the BiSb layer and the substrate. Indeed, XRD θ-2θ scans and rocking curves are shown in **(Figure S.5)** showing a similar pure $Bi_{0.85}Sb_{0.15}$ films grown along (0001) direction with an excellent crystallinity for both layer

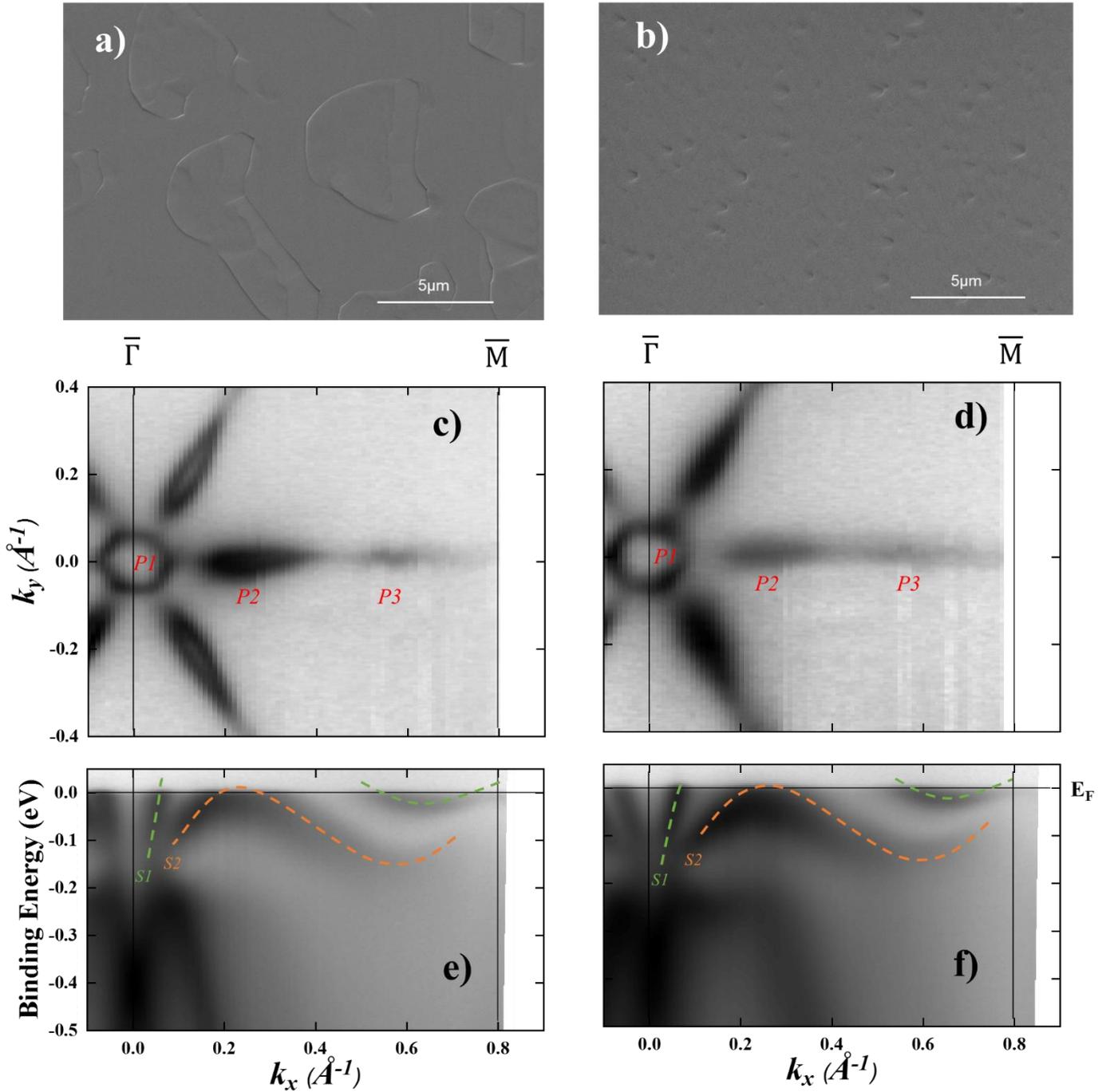

**Figure 5. a)** SEM image of the surface morphology of the 500nm thick $Bi_{0.85}Sb_{0.15}$ layer grown directly on GaAs (111)A substrate. **b)** SEM image of the surface morphology of a 500nm thick $Bi_{0.85}Sb_{0.15}$ layer grown in identical conditions but starting with an Sb bilayer on the GaAs (111)A surface. ARPES recorded on the same two layers: **c)** and **d)** Fermi surfaces recorded on conventional and domains-free films, respectively. **e)** and **f)** respective cross sections showing the band dispersion along the $\bar{\Gamma}\bar{M}$ direction of the same two samples.

Vertical black lines indicate the $\bar{\Gamma}$ and $\bar{M}$ points positions, respectively. Horizontal black lines indicate the Fermi level position. Green and orange dashed lines are guides for the eye.

The electronic structure evolution of both $Bi_{0.85}Sb_{0.15}$ films is measured by angular resolved photoemission spectroscopy (ARPES) to track any change in the topological surface states. Fermi surface maps obtained by measuring the band dispersion along $\bar{\Gamma}\bar{K}$ direction for conventional and domains-free films are shown in **(Figure 5.c.d)**, respectively. As previously reported,[11,29,30] the surface state intensity shows a hexagonal core (P1) centered at the $\bar{\Gamma}$ point and associated with the S1 surface states band crossing the Fermi level. The P1 hexagon is surrounded by six petals pointing in $\bar{\Gamma}\bar{M}$ directions, corresponding to the P2 hole pockets of the S2 band and P3 electron pockets of the S1 band local minima. We incidentally note here that despite the surface states display a sixfold symmetry for both films, the bulk bands' photoemitted intensity shows a strong threefold *C3* symmetry, reflecting the crystalline structure, as shown in **(Figure S.6)**. A change in the shape of the P2 petals is observed with a bands' broadening in the case of the domains-free film making difficult to judge the opening or closing of petals. **(Figure 5.e.f)** show the energy dispersion along the $\bar{\Gamma}\bar{M}$ direction, where the S1 and S2 surface states are illustrated respectively by green and orange dashed lines. In addition to the first crossing of S1 near $\bar{\Gamma}$, the S2 band exhibits two more Fermi level crossings for both films. A different shift can be observed for the domain-free film that might originate from a slight misalignment of the sample as shown in **(Figure S.7)**. Moreover, S1 band further double crosses the fermi level at the vicinity of $\bar{M}$ point, i.e. resulting in an odd number of Fermi level crossings for both samples, which is a direct signature of their topological nature and the presence of the TSS. We note that no S3 state is observed in our samples near the $\bar{M}$ point, as reported by Hsieh et *al.*.[11] Later theoretical and experimental studies demonstrate that this S3 band is induced by surface's imperfection and that only S1 and S2 surface bands should be observed.[29,31] The TSS of our samples are thus comparable to those recently reported for ultrathin BiSb layers and only achieved thanks to growth in a MBE chamber directly connected to the ARPES setup in an ultra-high vacuum environment.[29,32] These findings highlights the good crystalline quality of our samples as key parameter to perform ex-situ measurements and to directly probe topological properties. Combined with the wafer scale integration of domain-free TI on industrial substrates, our results should unlock new possibilities for future Spintronic devices.

## III. Conclusion

Topological insulator $Bi_{1-x}Sb_x$ thin films are successfully grown on GaAs (111)A industrial substrates by molecular beam epitaxy using two different protocols. Structural and electronic investigations are performed by making use of complementary and advanced tools. SEM observations indicate a composition- and thickness-dependant domains structure within the films. High-resolution XRD and TEM investigation reveal the true nature of such domains as an in-plane structural twining. Based on previous theoretical studies, the growth protocol is optimized with the insertion of an antimony bilayer prior to the TI growth. The optimized growth strategy is found to completely change the BiSb films' microstructure. Indeed, domains-free epitaxial layers are obtained with good crystallinity. ARPES measurements demonstrate that both growth processes lead to topologically protected surface state, with an odd Fermi level surface bands crossings. These results are in good agreement with recent studies reporting TSS on in-situ grown BiSb thin films. To the best of our knowledge, only textured or polycrystalline thin films have been grown at the wafer scale. We expect these findings to motivate additional efforts to hasten industrial integration of such material in future devices.

## 4. Experimental Section/Methods

*Material growth:* $Bi_{1-x}Sb_x$ thin films (x = [0.7; 0.11; 0.15; 0.19]) are grown on (111)A oriented GaAs substrates in a Riber MBE412 molecular beam epitaxy system. The substrate temperature is monitored thanks to a kSA BandiT system and the surface reconstruction is tracked during growth by reflection high energy electron diffraction (RHEED). The curvature of the sample is monitored in-situ during the growth using the Riber EZ-CURVE tool attached to the MBE reactor.[33]

*Structural characterization:* High-resolution X-Ray diffraction (HR-XRD) measurements using different geometries (θ-2θ, reciprocal space mapping, pole figure) are performed using a D8 Brucker diffractometer (λ = 1.54056 Å). Scanning electron microscopy (SEM) images are obtained using a FEI Aztec-600i microscope equipped with a Focused Ion Beam (FIB) cannon used for Transmission Electron microscopy (TEM) lamella preparation. TEM cross-sectional analysis is performed on 100nm-thick lamella using an ARM microscope by JEOL operated at 200kV, located at the Raimond Castaing Microanalysis Centre in Toulouse.

*ARPES measurements:* Angular resolved photo emission spectroscopy (ARPES) measurements are performed at CASSIOPEE beamline of the French national synchrotron facility SOLEIL. All measurements are performed at 17K with photon energies of 20eV.

ASSOCIATED CONTENT

Supporting information are available. SEM images analysis for films with different composition are supplied. Supplementary XRD and ARPES analysis are provided.

AUTHOR INFORMATION

**Corresponding Author**

**Mohamed Ali Khaled** − LAAS-CNRS, Université de Toulouse, Toulouse F-31400, France ; Email : mohamed-ali.khaled@laas.fr

**Sébastien R. Plissard** − LAAS-CNRS, Université de Toulouse, Toulouse F-31400, France

; Email : sebastien.plissard@laas.fr

**Author Contributions**

M.A.K., C.D. and S.R.P. conceived the project and experiments. A.A., S.F., Q.G., and S.R.P. grew the samples. S.R.P. and S.F. performed SEM observations. M.A.K. and S.F. performed the XRD measurements. L.C. performed the FIB preparation and measured the samples in the TEM. M.A.K., R.M, R.D, C.D., A.A., F.B., C.B., and S.R.P. performed ARPES samples preparation, measurements, and data analysis. M.A.K., C.D., and S.R.P. analysed the measurements and wrote the manuscript with inputs from all the authors.

**Funding Sources**

This work is founded by the Institute for Quantum Technologies in Occitanie under the grant name HyTop and by the Agence Nationale de la Recherche (ANR) under grant number ANR-17-PIRE-0001 (HYBRID project).

ACKNOWLEDGMENT

The authors thank Benjamin Reig for the help with TEM sample preparation. Authors acknowledge technical support provided by the LAAS-CNRS micro and nanotechnologies platform, member of French RENATECH network. This work was carried out in the EPICENTRE common laboratory between Riber and CNRS.

REFERENCES

1. Kane, C. L. & Mele, E. J. Z 2 Topological Order and the Quantum Spin Hall Effect. *Phys. Rev. Lett.* **95**, 146802 (2005).

2. Hasan, M. Z. & Kane, C. L. *Colloquium* : Topological insulators. *Rev. Mod. Phys.* **82**, 3045–3067 (2010).

3. Dankert, A., Geurs, J., Kamalakar, M. V., Charpentier, S. & Dash, S. P. Room Temperature Electrical Detection of Spin Polarized Currents in Topological Insulators. *Nano Lett.* **15**, 7976–7981 (2015).

4. Dankert, A. *et al.* Origin and evolution of surface spin current in topological insulators. *Phys. Rev. B* **97**, 125414 (2018).

5. Han, W., Otani, Y. & Maekawa, S. Quantum materials for spin and charge conversion. *Npj Quantum Mater.* **3**, 27 (2018).

6. Miron, I. M. *et al.* Perpendicular switching of a single ferromagnetic layer induced by in-plane current injection. *Nature* **476**, 189–193 (2011).

7. Manchon, A. *et al.* Current-induced spin-orbit torques in ferromagnetic and antiferromagnetic systems. *Rev. Mod. Phys.* **91**, 035004 (2019).

8. Liu, L., Moriyama, T., Ralph, D. C. & Buhrman, R. A. Spin-Torque Ferromagnetic Resonance Induced by the Spin Hall Effect. *Phys. Rev. Lett.* **106**, 036601 (2011).

9. Han, J. *et al.* Room-Temperature Spin-Orbit Torque Switching Induced by a Topological Insulator. *Phys. Rev. Lett.* **119**, 077702 (2017).

10. Lenoir, B. & Scherrert, S. TRANSPORT YT3PERTIES OF Bi-RICH Bi-Sb ALLOYS.

11. Hsieh, D. *et al.* A topological Dirac insulator in a quantum spin Hall phase. *Nature* **452**, 970–974 (2008).

12. Sasaki, M., Ohnishi, A., Das, N., Kim, K.-S. & Kim, H.-J. Observation of the possible chiral edge mode in $Bi_{1-x}Sb_x$. *New J. Phys.* **20**, 073038 (2018).

13. Gopal, R. K., Singh, S., Mandal, A., Sarkar, J. & Mitra, C. Topological delocalization and tuning of surface channel separation in Bi2Se2Te Topological Insulator Thin films. *Sci. Rep.* **7**, 4924 (2017).

14. Taskin, A. A., Ren, Z., Sasaki, S., Segawa, K. & Ando, Y. Observation of Dirac Holes and Electrons in a Topological Insulator. *Phys. Rev. Lett.* **107**, 016801 (2011).

15. Xu, Y. *et al.* Observation of topological surface state quantum Hall effect in an intrinsic three-dimensional topological insulator. *Nat. Phys.* **10**, 956–963 (2014).

16. Heremans, J. P., Cava, R. J. & Samarth, N. Tetradymites as thermoelectrics and topological insulators. *Nat. Rev. Mater.* **2**, 17049 (2017).

17. Khang, N. H. D., Ueda, Y. & Hai, P. N. A conductive topological insulator with large spin Hall effect for ultralow power spin–orbit torque switching. *Nat. Mater.* **17**, 808–813 (2018).

18. Lüpke, F. *et al.* In situ disentangling surface state transport channels of a topological insulator thin film by gating. *Npj Quantum Mater.* **3**, 46 (2018).

19. Yao, K., Khang, N. H. D. & Hai, P. N. Influence of crystal orientation and surface termination on the growth of BiSb thin films on GaAs substrates. *J. Cryst. Growth* **511**, 99–105 (2019).

20. Ueda, K., Hadate, Y., Suzuki, K. & Asano, H. Fabrication of high-quality epitaxial $Bi_{1-x}Sb_x$ films by two-step growth using molecular beam epitaxy. *Thin Solid Films* **713**, 138361 (2020).

21. Fan, T., Tobah, M., Shirokura, T., Huynh Duy Khang, N. & Nam Hai, P. Crystal growth and characterization of topological insulator BiSb thin films by sputtering deposition on sapphire substrates. *Jpn. J. Appl. Phys.* **59**, 063001 (2020).

22. Duy Khang, N. H. & Hai, P. N. Giant unidirectional spin Hall magnetoresistance in topological insulator – ferromagnetic semiconductor heterostructures. *J. Appl. Phys.* **126**, 233903 (2019).

23. Song, C. *et al.* Spin-orbit torques: Materials, mechanisms, performances, and potential applications. *Prog. Mater. Sci.* **118**, 100761 (2021).

24. Poh, H. Y. *et al.* Crystallinity Control of the Topological-Insulator Surface $Bi_{85}Sb_{15}$ ( 012 ) via Interfacial Engineering for Enhanced Spin-Orbit Torque. *Phys. Rev. Appl.* **19**, 034012 (2023).

25. Sadek, D. *et al.* Integration of the Rhombohedral BiSb(0001) Topological Insulator on a Cubic GaAs(001) Substrate. *ACS Appl. Mater. Interfaces* **13**, 36492–36498 (2021).

26. Sadek, D. *et al.* Structural and Electrical Characterizations of BiSb Topological Insulator Layers Epitaxially Integrated on GaAs. *Cryst. Growth Des.* **22**, 5081–5091 (2022).

27. Longo, E. *et al.* Exploiting the Close-to-Dirac Point Shift of the Fermi Level in the $Sb_2Te_3$/$Bi_2Te_3$ Topological Insulator Heterostructure for Spin-Charge Conversion. *ACS Appl. Mater. Interfaces* **15**, 50237–50245 (2023).

28. Sadek, D. *et al.* Growth of BiSb on GaAs (001) and (111)A surfaces: A joint experimental and theoretical study. *Appl. Surf. Sci.* **622**, 156688 (2023).

29. Benia, H. M., Straßer, C., Kern, K. & Ast, C. R. Surface band structure of $\text{Bi}_{1-x}\text{Sb}_{x}$(111). *Phys. Rev. B* **91**, 161406 (2015).

30. Nakamura, F. *et al.* Topological transition in Bi $1-x$ Sb $x$ studied as a function of Sb doping. *Phys. Rev. B* **84**, 235308 (2011).

31. Zhang, H.-J. *et al.* Electronic structures and surface states of the topological insulator Bi $1-x$ Sb $x$. *Phys. Rev. B* **80**, 085307 (2009).

32. Baringthon, L. *et al.* Topological surface states in ultrathin Bi $1-x$ Sb $x$ layers. *Phys. Rev. Mater.* **6**, 074204 (2022).

33. Arnoult, A. & Colin, J. Magnification inferred curvature for real-time curvature monitoring. *Sci. Rep.* **11**, 9393 (2021).

(34) S. Weber, "WinWulff." JCrystalSoft, 2018